\documentstyle[aps,epsfig]{revtex}  

\begin{document}

\draft

\title{Coexistence of pion condensation
and color superconductivity in two flavor quark matter}  
\author{M. Sadzikowski}   
\address{Institute of Nuclear Physics, Radzikowskiego 152, 
             31-342 Krak\'ow, Poland}    
\maketitle    

\tighten    

\begin{abstract} We show that the superconducting
2SC phase at high density and normal chiraly broken quark phase 
at low density is separated by the mixed non-uniform phase along the
baryon density line. 
\end{abstract}

\bigskip


\section{Introduction}  

In QCD with two flavors at high baryon density
the superconducting phase is the lowest energy state \cite{arw1,shur}. 
In this phase $u, d$ quarks
with two different colors (say "red", "green") create
Cooper pairs whereas the "blue" quarks remain free. The chiral
symmetry $SU(2)\times SU(2)$ is restored and only gauge color
group is broken. On the other hand at lower density one expects 
the state with broken chiral symmetry. This can be a hadronic
phase but the free quark plasma is not ruled out.
In such a situation the sequence of the phase transitions
along the density line (at zero or very low temperature)
follows the chain: hadronic phase/quark phase/superconducting
phase. All the phase transitions are expected to be 
first order. In this letter we present an explicit model describing the
transition between the normal quark and the superconducting phases.
It occurs that the transition actually goes through the mixed
state which consists of the non-uniform chiral phase and superconducting
phase. 

The non-uniform chiral phase had been introduced in the context of the non-relativistic
nuclear physics \cite{dautry} and then was extended for the systems of
high density plasma of quarks \cite{bron,sad1}. In this new phase
the chiral fields condense in both scalar and pseudoscalar channels 
creating a static chiral wave along one direction in space with
the wavelength $2\pi /|\vec{q}|$ where $\vec{q}$ is a wave vector.
It was shown that in some temperature - density range this configuration has lower energy
then the uniform quark phase. On the other hand we have learned from
the studies \cite{arw1,shur} that at higher densities one should
compare the non-uniform quark phase with the superconducting phase rather then with
the uniform quark phase. As a result one can show that the phase diagram
of strongly interacting matter is more complicated and contains a
region of the mixed state.

The high density matter exists in the Neutron Stars.
If the density is high enough then the superconducting phase
resides in their cores. In this situation
one can also expect the mixed phase to be present there -
the mixed state which separates uniform chiraly broken
phase from the superconducting phase.
We cannot prove that this mixed state is the same as
the one considered in this paper but we provide an explicit
realization of such a scenario. 

In the next section we introduce a thermodynamic potential
describing all the interesting phases. In the third section we present
the numerical analysis and finaly we discuss the results in the
Conclusions.

\section{Model of chiral and 2SC phases}

Let us consider the NJL model \cite{sad2}:
\begin{equation}
\label{h}
H=\int d^3x\left\{\bar{\psi}(-i\vec{\gamma}\cdot\vec{\nabla}-\mu\gamma_0)\psi
-G\left[ (\bar{\psi}\psi )^2+(\bar{\psi}i\gamma_5\vec{\tau}\psi )^2\right]
-G^\prime (\psi\tau_2t^AC\gamma_5\psi )^\dagger (\psi\tau_2t^AC\gamma_5\psi )\right\}
\end{equation}
where $\psi $ is the quark field, $\mu $ the quark chemical potential and
the color, flavor and spinor indices are suppressed. The vector
$\vec{\tau}$ is the isospin vector of Pauli matrices and $t^A$ are three color
antisymmetric group generators. The coupling constant
$G$ describes the interaction in the isospin singlet, Lorentz scalar and the isospin triplet,
Lorentz pseudoscalar, quark-antiquark channels whereas $G^\prime$ describes the interaction in
the color $\bar{\mbox{\bf{3}}}$, flavor singlet, Lorentz scalar diquark channel. 
Both couplings are related through the
Fierz transformation but we treat them as the independent parameters.
There is an additional parameter (or function) which regularize
infinities in the model. In our case we use the simplest momentum cut-off $\Lambda $.
Our main results are independent of this choice. 
We analize the four-fermion point interactions in the mean field
approximation using the general anstaz:
\begin{eqnarray}
\label{ansatz}
\langle\bar{\psi}\psi\rangle = -\frac{M}{2G}\cos\vec{q}\cdot\vec{x} \\\nonumber
\langle\bar{\psi}i\gamma_5\tau^a\psi\rangle = -\frac{M}{2G}\delta_{a3}\sin\vec{q}\cdot\vec{x} \\\nonumber
\langle\psi\tau_2t^AC\gamma_5\psi\rangle = \frac{\Delta}{2G^\prime}\delta_{A1} \\\nonumber
\end{eqnarray}
where the directions of symmetry breaking in the isospin and color spaces are
chosen arbitrarily. The wave vector $\vec{q}$ describes
the chiral wave along some arbitrary direction 
in space. In the limit of vanishing $\vec{q}$ we recover usual
uniform chiral phase. The gap parameter $\Delta $ describes the
superconducting phase. Let us notice that (\ref{ansatz}) is not
the analog of the LOFF phase in the superconductivity \cite{loff,sc_qcd} which arises due to
the difference in the chemical potentials of the pairing species. In our
case it is rather the dynamical interaction between the spin and isospin
degrees of freedom which effectively lower the energy \cite{bron,sad1}.

Using formulae (\ref{h}), (\ref{ansatz}) in the mean-field approximation we get:
\begin{equation}
\label{H_mf}
H_{MF}=\int d^3x\left\{\bar{\psi} (-i\vec{\gamma}\cdot\vec{\nabla}+MU-\mu\gamma_0)\psi
-\frac{\Delta}{2} (\psi\tau_2t^1C\gamma_5\psi )^\dagger 
-\frac{\Delta^\ast}{2}(\psi\tau_2t^1C\gamma_5\psi )
+\frac{M^2}{4G}+\frac{|\Delta|^2}{4G^\prime}\right\}
\end{equation}
where the Dirac operator is now space dependent
$U=\cos\vec{q}\cdot\vec{x}+i\gamma_5\tau_3\sin\vec{q}\cdot\vec{x}$. One
can eliminate this dependence by the unitary transformation to the
new fermi fields $\tilde{\psi} = U^{1/2}\psi $ which gives the free energy
in the form: 
\begin{equation}
\label{th_mf}
\tilde{H}_{MF}=\int d^3x\left\{\tilde{\psi}^\dagger 
(-i\vec{\alpha}\cdot\vec{\nabla}-\frac{1}{2}\vec{\Sigma}\cdot\vec{q}\tau_3+M\gamma_0-\mu )\tilde{\psi
}
-\frac{\Delta}{2} (\tilde{\psi}\tau_2t^1C\gamma_5\tilde{\psi})^\dagger 
-\frac{\Delta^\ast}{2}(\tilde{\psi}\tau_2t^1C\gamma_5\tilde{\psi})
+\frac{M^2}{4G}+\frac{|\Delta|^2}{4G^\prime}\right\}
\end{equation}
where $\vec{\Sigma}$ is the vector of Dirac spin matrices. The hamiltonian (\ref{th_mf})
describes two kinds of interactions: the "chiral" part and the "superconducting part". 
The chiral part mixes particles and antiparticles leaving color and 
isospin structure untouched. On the other hand the "superconducting" part mixes
colors and isospins of particles. To be more precise, one can introduce the plane wave basis
\begin{equation}\label{anni}
\tilde{\psi}^{j}_{\alpha }(t,\vec{x}) = \sum_{s=1,2}\int\frac{d^3k}{(2\pi )^3\sqrt{2 E(\vec{k})}}
\left\{ u_s(\vec{k}) a^j_{\alpha ,\, s}(\vec{k}) \exp (-i k x)+ 
v_s(\vec{k}) b^{j\,\dagger }_{\alpha ,\, s} (\vec{k}) \exp (i k x)\right\} ,
\end{equation}
where $E(\vec{k})=\sqrt{\vec{k}^2+M^2}$, $u_s, v_s$ are Dirac bispinors, $a^i_{\alpha ,\, s}(\vec{k})$ 
$(b^{i}_{\alpha ,\, s} (\vec{k}))$ is an
annihilation operator of quark (antiquark) of color $\alpha $, flavor $j$, spin $s$,
and momentum $\vec{k}$, satisfying usual anticommutation relations.
In this basis hamiltonian (\ref{th_mf}) creates the $24\times 24$ matrix which decays into
three independent diagonal $8\times 8$ blocks. Two of them are related to pairing between "red", "green" 
$u, d$ quarks and antiquarks whereas the third one describes the pairing between 
"blue" quarks and antiquarks (which do not create Cooper pairs). 
The hamiltonian (\ref{th_mf}) can be diagonalized with the positive energy eigenvalues
:
\begin{eqnarray}
\lambda_{1,\pm}(\vec{k}) = \sqrt{(E_\pm (\vec{k}) -\mu )^2+|\Delta |^2},\;\;\; 
\lambda_{2,\pm}(\vec{k}) = \sqrt{(E_\pm (\vec{k}) +\mu )^2+|\Delta |^2} \\\nonumber
E_\pm (\vec{k}) = \sqrt{\vec{k}^2+M^2+\frac{q^2}{4}\pm\sqrt{q^2M^2+\vec{k}\cdot\vec{q}}}
\end{eqnarray}
for $u, d$, "red", "green" quarks and $\lambda_{3,\pm}(\vec{k}) = E_{\pm }(\vec{k})$
for "blue" $u, d$ quarks. The negative eigenvalues has opposite signs. After 
diagonalization the hamiltonian (\ref{th_mf}) takes the general form $\tilde{H}_{MF} = \hat{H}+H_0$.
The first term is a diagonal operator and the second term defines
the energy of the ground state as a function of chemical potential. 
The vacuum expectation value $\langle\tilde{H}_{MF}\rangle = H_0$ and
\begin{eqnarray}
\label{h0}
H_0 = \frac{|\Delta |^2}{4G^\prime}+\frac{M^2}{4G}
-2\sum_{s=\pm}\int^\Lambda\frac{d^3k}{(2\pi )^3}(\lambda_{1,s}+\lambda_{2,s})
-2\sum_{s=\pm}\int^\Lambda\frac{d^3k}{(2\pi )^3}\lambda_{3,s}
+2\sum_{s=\pm}\int_{E_s<\mu}\frac{d^3k}{(2\pi )^3}(\lambda_{3,s}-\mu )
\end{eqnarray}
The first integral describes the contribution from the quarks which create
the chiral condensate as well as Cooper pairs. The factor 2 is the number of colors minus one
("red", "green"). The parameter $\Lambda $ reminds
of the momentum cut-off of the divergent integral. The last two terms are connected to
"blue" quarks which build only chiral condensate. The factor of 2, this time, describes
the number of flavors. Let us rewrite (\ref{h0}) in the form:
\begin{eqnarray}
\label{h02}
H_0 = \frac{|\Delta |^2}{4G^\prime}+\frac{M^2}{4G}
-6\sum_{s=\pm}\int^\Lambda\frac{d^3k}{(2\pi )^3}\lambda_{3,s}
-2\sum_{s=\pm}\left\{\int\frac{d^3k}{(2\pi )^3}(\lambda_{1,s}+\lambda_{2,s} - 2\lambda_{3,s})
+\sum_{s=\pm}\int_{E_s<\mu}\frac{d^3k}{(2\pi )^3}(\lambda_{3,s}-\mu )\right\}
\end{eqnarray}
which distinguishes explicitly between the vacuum contribution - the first integral, and 
the finite density contributions - the last two integrals. The vacuum contribution can be expanded
in the power of the wave vector $\vec{q}$ (nothing more but the gradient expansion in mesonic fields)
and we arrive at the final result:
\begin{eqnarray}
\label{h03}
H_{0}^{njl} = \frac{|\Delta |^2}{4G^\prime}+\frac{M^2}{4G}+\frac{M^2 F_{\pi}^2 \vec{q}^{\,2}}{2 M_{0}^2}
-12\int^\Lambda\frac{d^3k}{(2\pi )^3}E(\vec{k})
 \\\nonumber
-2\sum_{s=\pm}\left\{\int\frac{d^3k}{(2\pi )^3}(\lambda_{1,s}+\lambda_{2,s} - 2\lambda_{3,s})
-\int_{E_s<\mu}\frac{d^3k}{(2\pi )^3}(\lambda_{3,s}-\mu )\right\}
\end{eqnarray}
The coefficient at $\vec{q}^{\,2}$ is
model independent and it is related to the pion decay constant 
with its vacuum value $F_\pi = 93$ MeV. $M_0$ is the constituent quark mass at zero
density. Let us notice that in the limit 
of vanishing $|\Delta |$ we recover the chiral hamiltonian \cite{sad1}
whereas in the limit of vanishing $\vec{q}$ we can find the hamiltonian
describing Cooper pairing \cite{sad2}. Let us also mention that from the formula (\ref{h03})
one can easily guess the linear sigma model version of the hamiltonian. Indeed 
the
NJL and linear sigma models at the mean-field level are different only with respect to the description 
of the vacuum contribution to the energy. Thus instead of the formula (\ref{h03}) one can write 
in the linear sigma model:
\begin{eqnarray}
\label{h0ls}
H_{0}^{l\sigma} = \frac{|\Delta |^2}{4G^\prime}+\frac{M^2}{4G}+\frac{M^2 F_{\pi}^2 \vec{q}^2}{2 M_{0}^2}
+ \frac{m_{\sigma}^2}{8F_{\pi}^2}\left(\frac{m^2}{g^2}-F_{\pi}^2\right)^2
\\\nonumber
-2\sum_{s=\pm}\left\{\int\frac{d^3k}{(2\pi )^3}(\lambda_{1,s}+\lambda_{2,s} - 2\lambda_{3,s})
-\int_{E_s<\mu}\frac{d^3k}{(2\pi )^3}(\lambda_{3,s}-\mu )\right\}
\end{eqnarray}
where $g$ is dimensionless coupling constant and $m_\sigma$ is a mass of sigma meson.
However in the numerical calculation we only use the formula (\ref{h03}).

\section{Chiral/2SC phase transition}

The model parameters can be fitted to the values of the chiral
condensate and the pion decay constant which give
$G=5.01$ GeV$^{-2}$ and the cut-off $\Lambda = 0.65$ GeV \cite{klev}. For this set of parameters
the chiral phase transition at $\vec{q}=0$ and $|\Delta | = 0$
takes place at the chemical potential $\mu = 0.316$ GeV and the
value of the constituent quark mass at zero density is $M_0=0.301$ MeV. 
The coupling constant $G^\prime$ we treat as the additional
model parameter. The main results which are robust against the change
in $G^\prime$ are:
\begin{itemize}
\item there is the first order phase transition from the homogeneous chiraly broken
phase $M=M_0$, $\vec{q}=0$, $|\Delta |=0$ to the mixed phase
$M\neq M_0\neq 0$, $\vec{q}\neq 0$, $|\Delta |\neq 0$.
\item by increasing chemical potential one decreases constituent mass $M$
and increases $\vec{q}$ and $|\Delta |$.
\item there is additional first order phase transition from the
mixed phase to the 2SC phase ($M=0$, $\vec{q}=0$, $|\Delta |\neq 0$).
\end{itemize}
From the above points one can see that the 2SC phase is separated from
the chiraly broken phase by the the non-uniform mixed state. This conclusion
is still valid for non-zero temperatures. We discuss this feature
in the last section.
On the other hand the numerical details depend on the choice 
of $G^\prime$. In particular for the range $G/2\leq G^\prime\leq G$ we can find: 
\begin{itemize}
\item The transition to the mixed phase appears in the band of 10 MeV around $\mu = 0.29$ MeV
(smaller $G^\prime $ larger critical chemical potential)
.
\item The values of the gap parameter $|\Delta |$ can vary by a factor of two
(smaller $G^\prime $ larger $|\Delta |$).
\item The values of the wave vector $|\vec{q}|$ can change by 40-50 per cent
(smaller $G^\prime $ larger $|\vec{q} |$).
\item The range of the mixed phase along $\mu $-axis can change from 10 to 100 MeV
(smaller $G^\prime$ longer the range).
\end{itemize}

In the Fig. 1 we present the values of $M$, $|\vec{q}|$, $|\Delta |$ for $G^\prime = G/2$
as a function
 of chemical potential $\mu $. The first order phase transition takes place
at $\mu = 0.301$ GeV. The constituent mass drops to around half of its
vacuum value (Fig. 1a), the wave vector $|\vec{q}| = 0.34$ GeV (Fig. 1b) and
$|\Delta | = 0.004$ GeV (Fig. 1c). The mixed phase exists up to the second critical point $\mu = 0.396$ GeV
where another first order phase transition appears. The values of $M$, $\vec{q}$ drop
to zero and the value of the gap $|\Delta |=0.039$ GeV. For larger  $\mu$ 
we enter the homogeneous 2SC phase.

\begin{figure}
\centerline{\epsfxsize=5.5 cm \epsfbox{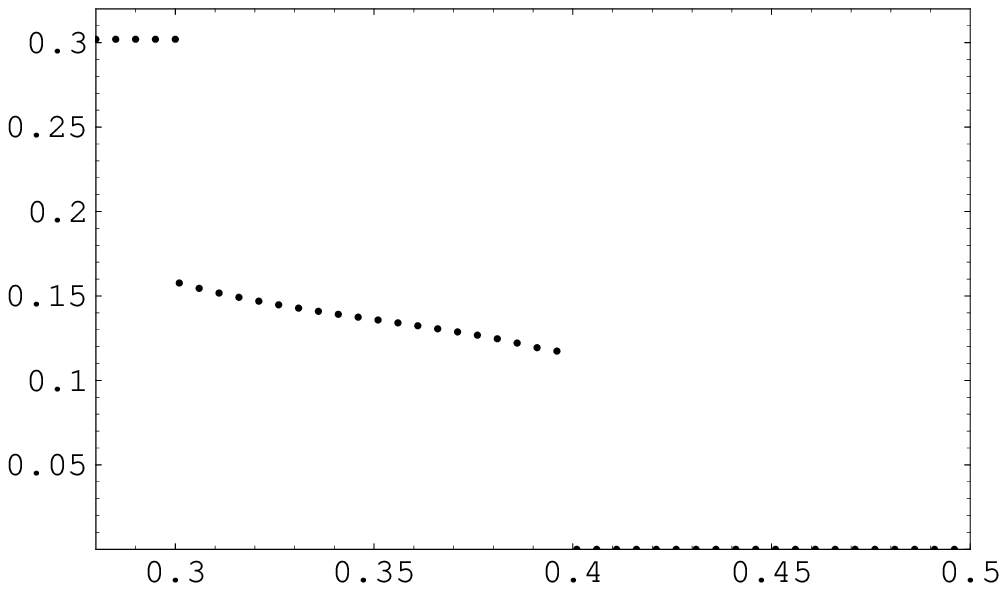} \epsfxsize=5.5 cm \epsfbox{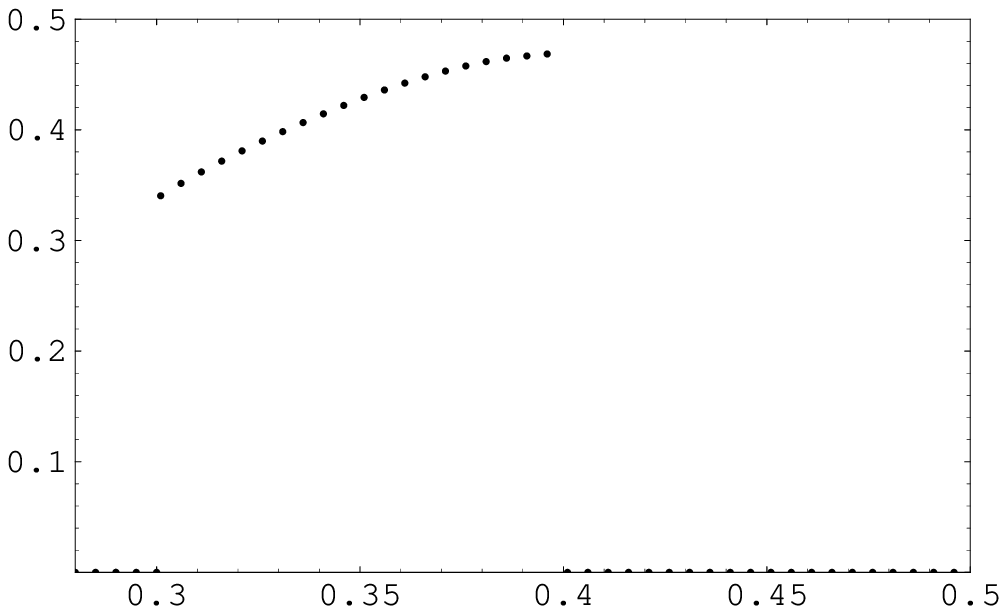} \epsfxsize=5.5 cm \epsfbox{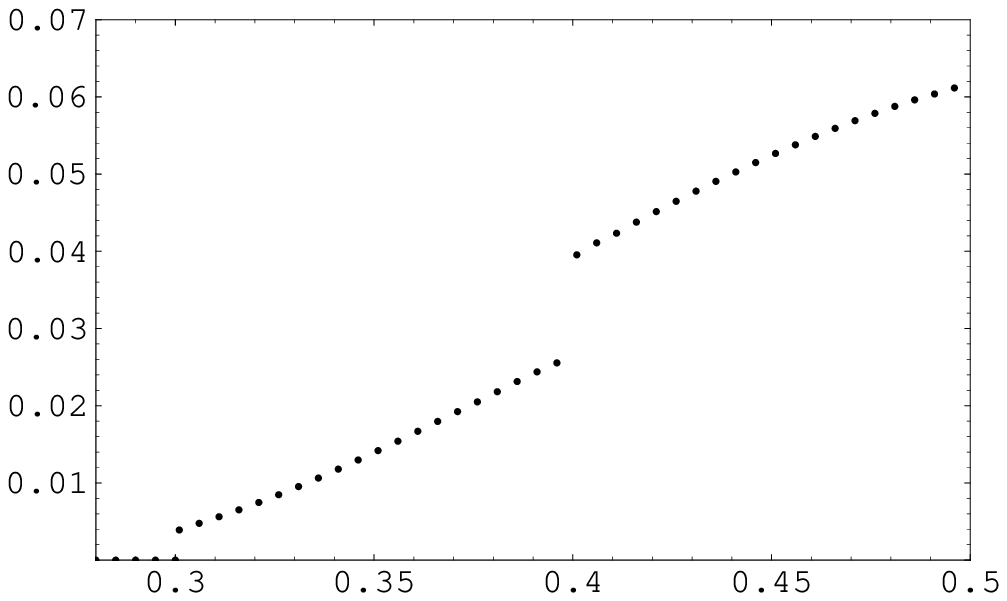}}
\caption{Dependence of $M$, $|\vec{q}|$ and $\Delta $ in GeV on the quark chemical potential $\mu $ in GeV.}
\end{figure}

\section{Conclusions}

In this letter we showed that the high density superconducting phase 2SC
is separated from the low density chiraly broken phase by the region of
the mixed non-homogeneous phase. This mixed state consists of the chiral waves
with non-zero expectation values in the scalar and pseudoscalar channels
$\langle\bar{\psi}\psi\rangle ,\langle\bar{\psi}i\gamma_5\tau_a\psi\rangle $ 
described by the wave vector $\vec{q}$ which points at arbitrary direction is space.
In addition there is a non-zero value of the superconducting gap parameter 
$|\Delta |$. We cannot prove that this mixed state is the lowest energy
state however we showed that its energy is lower then that of the homogeneous phases.
Similar questions was already addressed in the context of
the Overhauser effect \cite{deryagin} versus BCS pairing. It was shown
that both types of pairing can be competitive \cite{park,rapp} with
each other. However
it was also shown that within the NJL-type of models
BCS state is favoured over the Overhauser state \cite{rapp} which suggests that
the non-homogeneous phase (\ref{ansatz}) is
lower then BCS and Overhauser phases in the considered density
range. Nevertheless one has to keep in mind that other approaches
(instanton models) favours Overhauser effect over BCS pairing
\cite{park,rapp} thus one can not drew the final conclusion and more work is
required.
We can then conclude that the superconducting phase and the normal phase are
separated by some kind of crystal state. Let us stress that the crystal 
structure (\ref{ansatz}) is dynamical in origin, not "kinematical" in a sense of the LOFF state
\cite{loff,sc_qcd}. 

Our results are robust against the reasonable changes in the model parameters.
What is more the same results one can derive not only in the NJL model but
also in the linear sigma model. Thus also in that sense the results are model independent.
The numerical details, like the location of the critical points, the values of the constituent
mass, wave vector and superconducting gap, depends on the choice of the parameters.
Although we perform our calculations for massless quarks the introduction of the small
current quark mass does not change the main conclusions. It merely distorts the shape
of the chiral wave as was shown in \cite{thies} for one dimensional system.
One can also expect that the mixed phase exists for 
some range of temperatures (below $T\sim 50$ MeV where superconductor
melts) and densities. However the detailed picture of
the phase diagram remain to be done.

The real QCD consists of two light flavors and one heavier, strange quark. In this
situation one has to consider the three flavors NJL model. 
However we expect that our main points remain the same. 
If one considers the superconducting phase inside the Neutron Stars then 
the question of the color and charge neutrality rises \cite{alford}.
It was shown by the explicit calculation \cite{reddy} that at non-zero lepton chemical
potential the 2SC phase borders with the normal quark phase at lower densities and
only at higher densities it changes into the CFL phase. Thus even in the case of the three flavors 
QCD we can still have quark matter/2SC
boundary in the Neutron Stars. This boundary is not just a line in the phase
diagram but the mixed state region as we have shown.
What kind of the crystal 
or non-uniformity is realized in the three flavors QCD has not been discoverd so far.

{\bf Acknowledgement} 
This work was supported by a fellowship from the Foundation for Polish Science.
It was also supported in part by Polish State Committee for Scientific Research, 
grant no. 2P 03B 094 19.

\end{document}